\documentclass[apex, onecolumn]{jjap3_arXiv}
\usepackage{txfonts}
\usepackage{braket}
\title{Electric-field-induced $Z_{2}$ topological phase transition in strained 
single bilayer Bi(111)}%
\author{Hikaru Sawahata$^{1}$\thanks{E-mail: sawahata@cphys.s.kanazawa-u.ac.jp}, Naoya Yamaguchi$^1$, 
Fumiyuki Ishii$^2$\thanks{E-mail: ishii@cphys.s.kanazawa-u.ac.jp}}
\inst{
$^1$
Graduate School of Natural Science and Technology, Kanazawa University, Kanazawa 920-1192, Japan
\\
$^2$
Nanomaterials Research Institute, Kanazawa University, Kanazawa 920-1192, Japan
}
\abst{
For controlling the critical electric fields of the topological phase transition in single bilayer Bi(111), we investigated topological phases in a strained system through first-principles calculations. We found a quadratic band touching semimetallic state at tensile strain $\epsilon=0.5$\%. Around this strain, the topological phase can be switched to a trivial insulator by an infinitesimal electric field. The positions at which Dirac cones appear in the electric-field-induced topological phase transition changed for the strain $\epsilon>0.5$\% and $\epsilon<0.5$\%. Our results indicate that this topological phase transition could be applied to novel spintronic devices.
}
\begin{document}
\maketitle
The electric-field-driven $Z_2$ topological phase transition plays an important role in the application of topological materials\cite{Hasan_Colloquium_2010, Ando_Topological_2013} to novel devices. The $Z_2$ topological phase has dissipation-free spin currents at the edge of the system, and its special edge state is robust against nonmagnetic impurities. If the topological phases could be switched by electric fields, novel spintronic devices using edge spin currents could be realized\cite{Liu_Switching_2015}.
\par
A Bi(111) thin-film is a candidate material for fabricating an electric-field-switched $Z_2$ topological phase device. It is reported that Bi(111) thin-film can be formed on Si(111)\cite{Nagao_Nanofilm_2004}. Bi(111) thin-film and single bilayer Bi(111) are an important material in two-dimensional topological insulators so that it has been extensively studied theoretically and experimentally\cite{Murakami_Quantum_2006,Yaginuma_Electronic_2008,Wada_Localized_2011,Hirahara_Interfacing_2011,Yao_Topologically_2016}. 
We previously predicted that single bilayer Bi(111) is a topological insulator under electric field of $E<2.1$ V/{\AA} and a trivial insulator under $E>2.1$ V/{\AA} by computing $Z_2$ invariants\cite{Sawahata_First_2018} and edge states\cite{Sawahata_Electric_2018}. The bandgap of single bilayer Bi(111) decreases from 0.32 eV to 0 eV when the applied electric field reaches $E=2.1$ V/{\AA}. 
To realize the device applications using single bilayer Bi(111), this critical electric field $E=2.1$ V/{\AA} is too large so that we should reduce the critical electric field by tuning the bandgap.
\par
The epitaxial strain can tune the bandgap of single bilayer Bi(111).
Density functional calculations predicted that 
the bandgap is changed and closed by the tensile strain
\cite{Chen_Robustness_2013,Huang_Strain_2014,
Wang_Topological_2017}.
In the previous experiment, Bi(111) thin-film formed on a Si(111) and Bi$_{2}$Te$_{3}$ substrate shows the lattice constant $a=4.54$ {\AA}\cite{Nagao_Nanofilm_2004} and $a=4.38$ {\AA}\cite{Hirahara_Interfacing_2011} respectively.
It is reported that
topological electronic states of Bi(111) thin-films are changed
by the epitaxial strain of these different substrate.
\cite{Hirahara_Atomic_2012}.
\par
In this study, based on first-principles calculations, we demonstrated that the critical electric field in single bilayer Bi(111) can be reduced by applying a small tensile strain $\epsilon=0.5\%$. 
First, we investigated the bandgap in strained systems. We found a critical strain of bandgap closing and a quadratic band touching semimetallic state.
We computed $Z_2$ invariants of two different insulator phases that appear under strain and confirmed that both are $Z_{2}$ topological insulator phases. 
We also computed the $Z_2$ topological phase in strained systems under applied electric fields. Unlike in our previous studies, we found that the topological phase transition was induced by small electric fields for tensile strain of $\epsilon\simeq0.5$\%. 
The positions at which Dirac cones appear in the electric-field-induced topological phase transition changed for the strain $\epsilon>0.5$\% and $\epsilon<0.5$\%.
\par
\begin{figure}[htbp]
\begin{center}
    \includegraphics[width=\columnwidth]{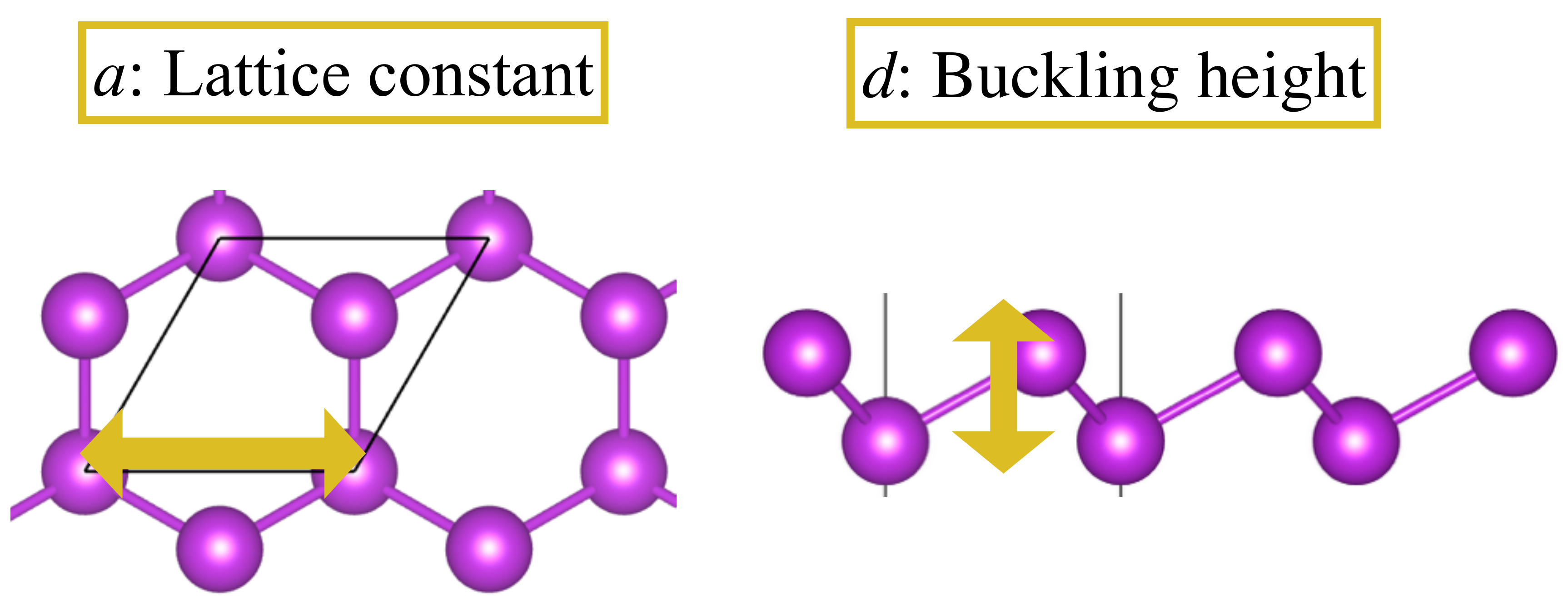}
    \end{center}
    \caption{Structure of single bilayer Bi(111) in a film. We optimized the buckling height with a strained lattice constant.}
    \label{parameterfig}
\end{figure}

Figure \ref{parameterfig} shows the structure of single bilayer Bi(111). 
Two atoms in the unit cell of a hexagonal lattice are set. 
We optimized the buckling height $d$ with the strained lattice constant $a$. 
We define the tensile strain as $\epsilon=(a-a_{\rm exp})/a_{\rm exp}$, 
where $a_{\rm exp} = 4.54$ {\AA} is an experimental lattice constant 
and the buckling height is $d_{\rm exp}=1.45$ {\AA}\cite{Nagao_Nanofilm_2004}. 
This strain changes the buckling height.
The bandgap of other hexagonal lattice systems like a graphene and silicene
can be tuned by the changing buckling height\cite{Ezawa_Monolayer_2015}. Therefore, we expect that the bandgap of single bilayer Bi(111) is related to the buckling height $d$.
\par
We performed density functional calculations using OpenMX code\cite{openmx}. 
We used the local spin density approximation\cite{Ceperley_Ground_1980,Perdew_Self_1981} as the exchange correlation functional. 
We used norm-conserving pseudopotentials\cite{Hamann_Norm_1979} and the linear combination of multiple pseudoatomic orbitals\cite{Ozaki_Variationally_2003,Ozaki_Numerical_2004} for wave function expansion. 
We set the pseudoatomic basis as Bi8.0-s3p3d2; this indicates a cutoff radius of 8.0 Bohr and pseudoatomic orbitals as three s-orbitals, three p-orbitals, and two d-orbitals. 
Spin-orbit interactions were included by a $j$-dependent pseudopotential composed relativistically (fully relativistic pseudopotential)\cite{Theurich_Self_2001}, where $j$ is the total angular momentum. 
We set ${\bf k}$-space sampling points of $13\times13\times1$ for reciprocal lattice vectors and cutoff energy of 300 Ry. 
Electric fields were introduced as sawtooth potentials\cite{Kunc_External_1983,Resta_Self_1986} and we assumed that the lattice parameters and atomic positions were not changed by electric fields.
\par
For confirming $Z_2$ topological phases 
of the system, 
we used 
the lattice Chern number method\cite{Fukui_Quantum_2007,Feng_First_2012}.
The $Z_2$ topological phase is determined by a $Z_2$ invariant\cite{Fu_Time_2006}. $Z_2 = 1$ corresponds to a topological insulator phase, and $Z_2 = 0$ corresponds to a trivial insulator phase. 
The lattice Chern number method computes the $Z_2$ invariant as $Z_2 = 1/2\pi i(\int{\bf A}\cdot d{\bf k}-\int F_z dk_x dk_y)$ (mod\ 2) on a half Brillouin zone $[-{\bf G}_1/2, {\bf G}_1/2]\otimes[0, {\bf G}_{2}/2]$, where ${\bf A}_{n} = \bra{u_{n{\bf k}}} \partial_{\bf k} \ket{u_{n{\bf k}}}$ is called the Berry connection and ${\bf F}=\nabla\times{\bf A}$ is called the Berry curvature. 
This method can be applied to the system with the broken spatial inversion symmetry.
We also use the parity method\cite{Fu_Topological_2007}  for computing $Z_{2}$ invariant of the system without electric fields.
The parity method computes the $Z_2$ invariant as $\prod_{i=1}^4\delta_{i} = (-1)^{Z_2}$, where $\delta_i\ (i = 1, 2, 3, 4)$ is the parity sign on ${\bf k}$ time-reversal-invariant points ($\Gamma_{1}: 2\pi/a(0, 0, 0), \Gamma_{2}: 2\pi/a(0.5, 0, 0), \Gamma_{3}: 2\pi/a(0, 0.5, 0)$, and $\Gamma_{4}: 2\pi/a(0.5, 0.5, 0)$). Here, ${\bf k} = 2\pi/a(k_{1}, k_{2}, k_{3})$ implies that ${\bf k} = k_1{\bf G}_1 + k_2{\bf G}_2 + k_3{\bf G}_3$, where ${\bf G}_m\ (m = 1, 2, 3)$ is a reciprocal lattice vector. 
We implemented these two methods in OpenMX. 
\par
First, we optimized the buckling height when the lattice constant is strained as $-5\% <\epsilon<+5\%$ without an electric field. The buckling height decreases as the lattice constant increases. It is $d=1.75$ {\AA} at $\epsilon=-5\%$ and $d=1.47$ {\AA} at $\epsilon=+5\%$. When we optimize the atomic structure at $a_{\rm exp}$, the obtained buckling height is 1.63 {\AA}, larger than the experimental parameter $d_{\rm exp}=1.45$ {\AA}\cite{Nagao_Nanofilm_2004}. This is because our calculated system has single bilayer, whereas seven or more Bi(111) bilayers were formed on Si(111) experimentally.
\par
\begin{figure}[htbp]
    \begin{center}
    \includegraphics[width=.8\columnwidth]{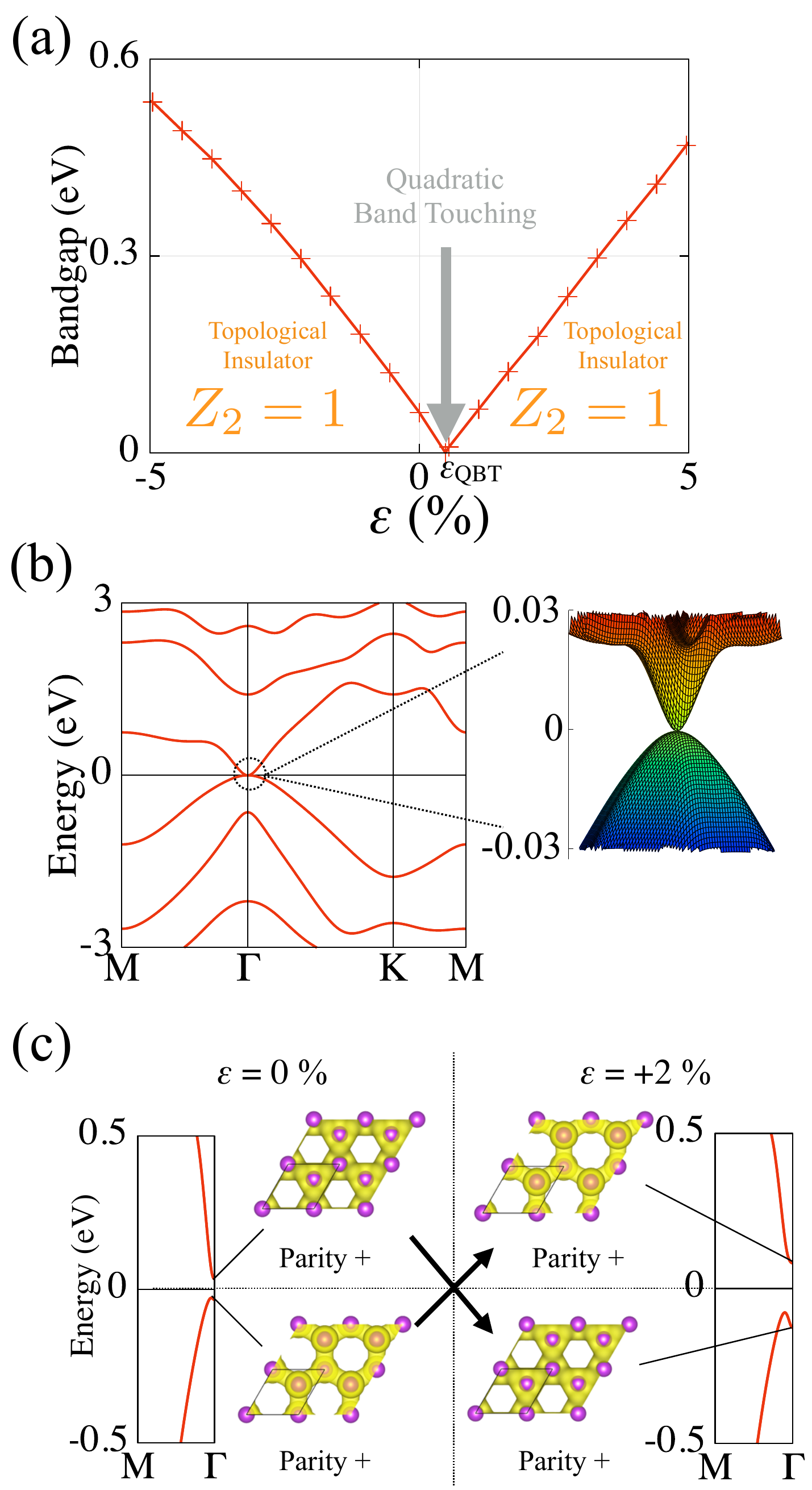}
    \end{center}
    \caption{(a) Strain dependence of bandgap. (b) Band dispersion at $\epsilon=\epsilon_{\rm QBT} = +0.5 \%$, enlarged view: quadratic band touching at $\Gamma$ point. (c) Band inversion at $\Gamma$ point, parity sign, and wave functions.}
    \label{StrainPhase}
\end{figure}

Next, we investigated the band structure for the strain of $-5\% < \epsilon < +5\%$ without electric fields. 
Figure \ref{StrainPhase}(a) shows the bandgap for various strains 
without electric fields. 
For $\epsilon=-5\%$, the bandgap is 0.53 eV and it 
decreases monotonically with increasing strain. 
This bandgap decreasing is originated from change in 
buckling height\cite{Ezawa_Monolayer_2015},
$d=1.75$ {\AA} for $\epsilon=-5\%$ and $d=1.61$ {\AA} 
for $\epsilon=0.5\%$.
At $\epsilon=\epsilon_{\rm QBT}=0.5\%$, 
the bandgap is closed at the $\Gamma$ point 
by the spin-orbit interaction, 
and the system shows a semimetallic quadratic band touching (QBT) state\cite{Chong_Effective_2008,Sun_Topological_2009}, 
as shown in Fig. \ref{StrainPhase}(b). 
Upon further increasing the strain, 
the bandgap is opened and reaches 0.47 eV at $\epsilon=+5\%$. 
This bandgap increasing can be understood by changing spin-orbit coupling
induced by localization of wave functions.
We confirmed the mechanism of the bandgap by calculating band structures without spin-orbit coupling.
On turning off spin-orbit coupling, 
the bandgap is opened for $\epsilon<\epsilon_{\rm QBT}$, while the bandgap is closed for $\epsilon>\epsilon_{\rm QBT}$.
\par
We calculated $Z_2$ invariants by the parity method and lattice Chern number method. 
The strain-induced $Z_{2}$ topological 
phase transition occurs when the bandgap is 
closing\cite{Liu_Strain_2016}. 
However, the system always has $Z_2=1$; therefore, strained 
single bilayer Bi(111) preserves the $Z_{2}$ 
topological insulator phase\cite{Huang_Strain_2014,Wang_Topological_2017}. 
Figure \ref{StrainPhase}(c) shows the wave function and parity sign at the $\Gamma$ point at around the critical strain. We can confirm band inversion at the $\Gamma$ point by investigating the shapes of the wave functions on the valence band and conduction band; however, the parity sign of both bands is equivalent, and therefore, the result of computing $Z_2$ invariants does not change. 
This result may indicate that another topological character changes by the strain instead of $Z_{2}$ invariant.
\par
\begin{figure}[htbp]
\begin{center}
    \includegraphics[width=\columnwidth]{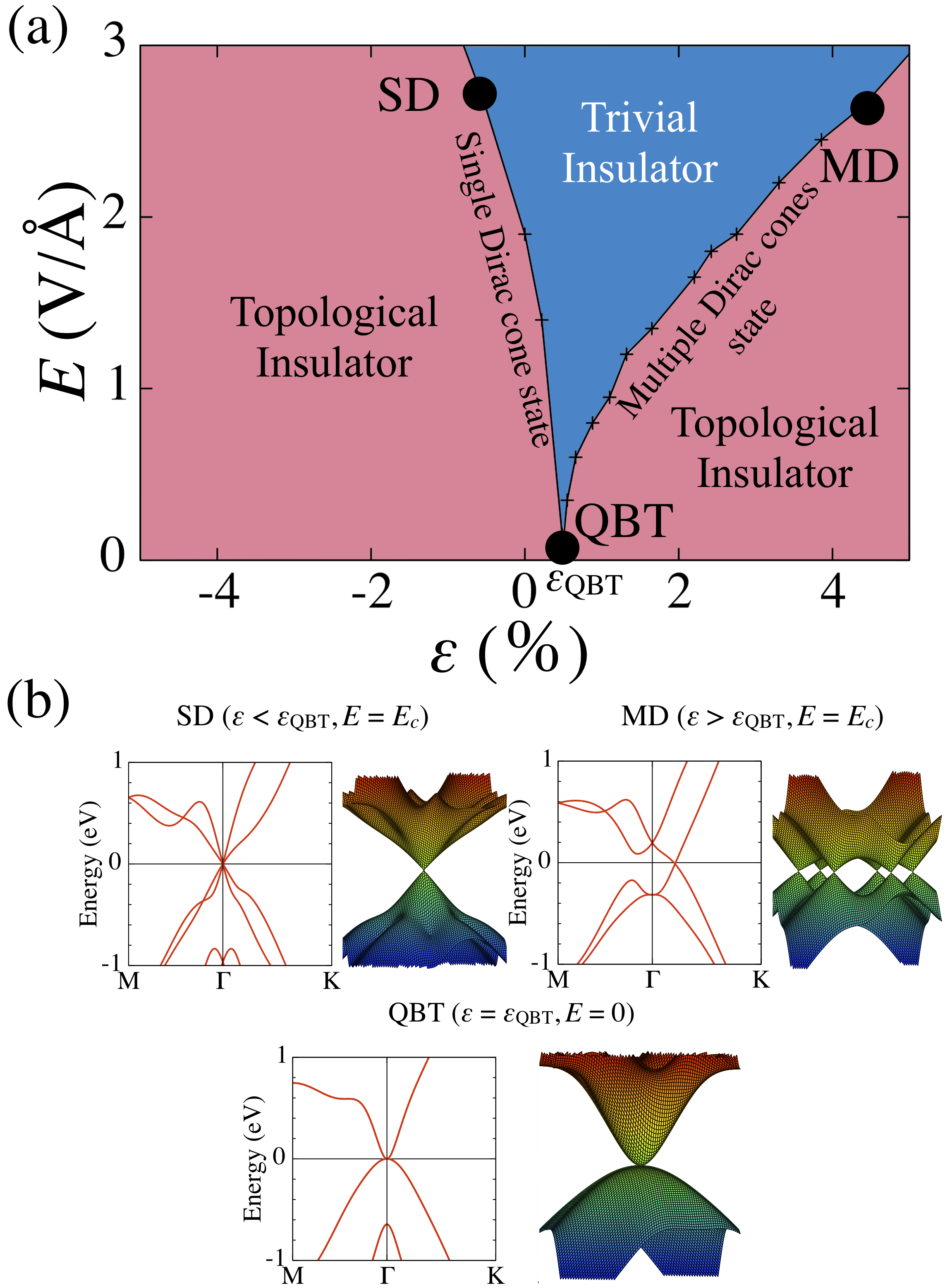}
    \end{center}
    \caption{(a) Topological phase diagram and critical electric field (data points) for various strains. (b) Band structure at critical electric field for various strains. 
    The example of single Dirac cone state (SD) is 
    the system at $\epsilon=-0.6$\% and $E=2.6$ V/{\AA}.
    The example of multiple Dirac cones state (MD) is 
    the system at $\epsilon=4.4$\% and $E=2.6$ V/{\AA}.
    Positions of Dirac cones changed for $\epsilon < \epsilon_{\rm QBT}$ and $\epsilon>\epsilon_{\rm QBT}$.}
    \label{Diracpoint}
\end{figure}

We investigated the electric-field-induced $Z_{2}$ topological phase transition.
Figure \ref{Diracpoint}(a) shows a topological phase diagram of strained systems under electric fields. 
The topological insulator phases are switched to trivial insulator phases by electric fields, as reported in our previous study
\cite{Sawahata_First_2018}. 
For $\epsilon<\epsilon_{\rm QBT}$, the critical electric field is drastically enhanced by compressive strain compared to tensile strain for $\epsilon>\epsilon_{\rm QBT}$.
An important result is the fact that the topological phase of the system around $\epsilon=\epsilon_{\rm QBT}$ can be switched by an infinitesimal electric field; therefore, we can achieve topological phase switching by realistic electric fields if we use these strain states.
\par
The Dirac cones appeared in the electric-field-induced $Z_{2}$ topological phase transition are different 
between $\epsilon>\epsilon_{\rm QBT}$ and $\epsilon<\epsilon_{\rm QBT}$. 
When the $Z_2$ topological phase changes in the two-dimensional system, Dirac semimetals appear\cite{Murakami_Phase_2007}.
Figure \ref{Diracpoint}(b) shows the band structure at the critical electric field for various strains and at the critical tensile strain.
For $\epsilon<\epsilon_{\rm QBT}$,  Dirac cones appear at the $\Gamma$ point (single Dirac cone (SD) state in the phase diagram Fig. \ref{Diracpoint}(a)) when the bandgap closes under the applied electric field. 
Two unoccupied bands and two valence bands are degenerate at the 
$\Gamma$ point in the SD state.
However, for $\epsilon>\epsilon_{\rm QBT}$, six Dirac cones appear on the $\Gamma$-K line (multiple Dirac cone (MD) state in the phase diagram Fig. \ref{Diracpoint}(a)) as in our previous study\cite{Sawahata_First_2018}. 
One unoccupied band and one valence band are degenerate on the every Dirac cone in the MD state.
%
The appearance of these Dirac cones are closely linked to the QBT state.
The QBT is parabolically band crossing at the Fermi energy
\cite{Chong_Effective_2008,Sun_Topological_2009} 
and its Berry flux 
$\Phi=-i\int d{\bf k}\cdot {\bf A}$ is 0 
at this point in the case of accidental band crossing.
The QBT in the present system appears at $\Gamma$ point (QBT in the phase diagram Fig. \ref{Diracpoint}(a)), two unoccupied bands and two valence bands are degenerate.
This QBT can split into several Dirac cones with $\Phi=\pm\pi$ 
with broken spatial inversion symmetry
while preserving the total Berry flux $\Phi=0$.
In the SD state, we confirm $\Phi=0$ at $\Gamma$ point because Dirac cones are degenerate and both Berry flux $\Phi=\pm\pi$ of Dirac cones cancel each other out, but its band dispersion is linear clearly.
In the MD state, we confirm $\Phi=-\pi$ for three Dirac cones and $\Phi=\pi$ for three Dirac cones\cite{doi:10.1143/JPSJ.74.1674}.
\par
In summary, we investigated the strains and electric field effects of single bilayer Bi(111) through a first-principles study. 
We computed the bandgap and topological phase against the strain and found that the bandgap was closed and 
the quadratic band touching semimetallic state appeared
at the $\Gamma$ point for $\epsilon=\epsilon_{\rm QBT}$; however, the $Z_2$ topological phase did not change. 
We also investigated the $Z_2$ topological phase under applied electric fields. 
The topological phase can be switched by an infinitesimal electric field near $\epsilon=\epsilon_{\rm QBT}$, and we achieve switching of the $Z_2$ topological phase of single bilayer Bi(111) by reasonable electric fields.
For $\epsilon>\epsilon_{\rm QBT}$, the electric-field-induced topological phase transition occurs as described in our previous study\cite{Sawahata_First_2018}; six Dirac cones appear on the $\Gamma$-K line. 
On the other hand, for $\epsilon<\epsilon_{\rm QBT}$, the topological phase transition occurs differently; degenerate Dirac cones appear at the $\Gamma$ point. This difference may be related to the experimental study of Bi(111) thin-film showing different surface states under a tensile strain\cite{Hirahara_Atomic_2012}.
\acknowledgment
This work was supported by a Grant-in-Aid for Scientific Research on Innovative Area "Nano Spin Conversion Science" (Grant No. 17H05180). This work was also supported by a JSPS Grant-in-Aid for Scientific Research on Innovative Areas "Discrete Geometric Analysis for Materials Design" (Grant No. 18H04481). This work was partially supported by Grants-in-Aid on Scientific Research under Grant No. 16K04875 from the Japan Society for the Promotion of Science and Master 21 from the Yoshida scholarship foundation. The computations in this research were performed using the supercomputers at RIIT, Kyushu University, and the ISSP, University of Tokyo.
\bibliographystyle{jjap}
\providecommand{\noopsort}[1]{}\providecommand{\singleletter}[1]{#1}%

\end{document}